\documentclass[twocolumn,showpacs,preprintnumbers,amsmath,amssymb]{revtex4}
\usepackage{graphicx}% Include figure files
\usepackage{dcolumn}% Align table columns on decimal point
\usepackage{bm}% bold math

\begin{document}

\title{Suppression of Complex Spiral--Wave Activity in an Ionic Model
of Cardiac Tissue by Weak Local Stimulations}

\author{Ekaterina Zhuchkova}
\email{zhkatya@polly.phys.msu.ru}
\author{Boris Radnaev}
\email{radnaev@polly.phys.msu.ru}
\author{Alexander Loskutov}
\email{loskutov@chaos.phys.msu.ru}
\thanks{corresponding author}
 \affiliation{
M.V. Lomonosov Moscow State University, Physics Faculty, 119992 Moscow, Russia}

\date{\today}

\begin{abstract}

On the basis of a quite realistic ionic Fenton--Karma model of the
cardiac tissue we consider the problem of defibrillation by a
local weak forcing. In contrast to other systems, this model
accurately reproduces the most essential mesoscopic properties of
the cardiac activity. It is shown that suppression of spiral--wave
turbulent dynamics in the heart tissue may be realized by a
low-voltage local non-feedback electrical stimulation of
monophasic and biphasic shapes. After stabilization the medium
goes to a spatially homogeneous steady state.

\end{abstract}

\maketitle

Cardiovascular diseases (CVDs) are responsible for more than 4 million deaths
each year in Europe (over 1.5 million deaths in the EU) and account for about
30\% of life-span loss in Europe (over 30\% in the EU) \cite{Stat}. A major group
of CVDs involves disturbances of the normal cardiac rhythm (cardiac arrhythmias).
The extreme form of cardiac arrhythmias and the prevalent mode of the sudden
death among patients with CVDs is ventricular fibrillation (VF), which is a fast
developing disturbance of spatially organized contraction of ventricles that is a
consequence of abnormalities of electrical conduction in the heart muscle.
Following to the contemporary conjecture, VF is produced by a multiple wavelet
re-entry, which is spiral waves in 2D and scroll waves in 3D \cite{ZJ} (i.e. by
spatio--temporal chaos or spiral--wave turbulence).

Since VF, sustained for only a few minutes, leads to death, an immediate
intervention is required. In emergency care medicine the application of
high--energy electrical stimulation through the patient's chest is commonly used
to suppress the fibrillation and restore the normal rhythmicity of the heartbeat.
However, high--energy shock can cause the necrosis of myocardium or give rise to
functional damage manifested as disturbances in atrioventricular conduction.

The application of electrical pulses for the termination of fibrillations is also
used in implantable cardioverter defibrillators (ICDs). These devices surgically
implanted into the bodies of high--risk cardiac patients and initiating
low--power electrical pacing pulses automatically when they detect a dangerous
activity. However, in the case of complex arrhythmias (but not fibrillation),
some patients may also undergo the ICD action. As a consequence, such patients
additionally have a severe pain. Therefore, a very important factor in the design
of modern ICDs is decreasing the stimulation amplitude in order to avoid a
painful high energy shock and damage to the heart itself and surrounding tissues.
Thus, there is high demand in clinics on alternative methods of defibrillation
which would work with lower voltages.

The recent research \cite{takagietal} may provide an alternative to the
conventional ICD therapy by terminating re-entrant arrhythmias with the field
strengths that are 5 to 10 times (or delivered energy 25-100 times) weaker than
usual defibrillation shocks. However, the method used in \cite{takagietal} is
valid only for the high--risk cardiac patients who had previous myocardial
infarctions (``heart attacks'').

Theoretical studies suggest that low-energy defibrillation protocols are also
possible at exploiting the dynamical properties of re-entrant waves under
electrical forcing, known as feedback-driven resonant drift \cite{biktashev}.
However, the major problem in practical use of the resonant drift is the change
of the resonant frequency with the position of rotating wave, in particular,
close to unexcitable boundaries.

The qualitatively different approach of low-amplitude suppression the complex
dynamics of nonlinear systems by application of perturbations without feedback
was firstly proposed in \cite{alos} and mathematically substantiated in
\cite{loshi}. Recently this method was tested on a model of the
Belousov-Zhabotinsky reaction and showed its validity \cite{kovlos}. The same
parametric controlling but involving the feedback was used in \cite{alsam}.
Although parametric (non-feedback) suppression or (feedback) controlling leads to
the stabilization of complex dynamics, it is not realizable for the electrical
defibrillation. The only application seems to be in formulating drug therapies
which modify ionic currents in order, for example, to prevent alternans
(beat-to-beat alternation in the action potential duration) which is presumably
one of the causes of breakup of a single rotating wave into multiple re-entrant
waves (see, e.g. \cite{echekar, stoscol, allexot}).

The recent investigations of the active medium theory offer new opportunities for
the electrical defibrillation: The amplitude of the external stimulation can be
\textit{essentially} decreased and the turbulent regime in excitable systems may
be suppressed by a sufficiently weak periodic external forcing applied globally
\cite{gray} or locally \cite{LCV, VCL, LVys, Zhang}. By this manner, it is
possible not only to suppress spatio--temporal chaos and stabilize the media
dynamics, but also reestablish the initial cardiac rhythm, because after
stabilization the medium goes to a spatially homogeneous steady state.

In the present paper, on the basis of a realistic cardiac model we resolve the
problem of suppressing the fibrillative activity by a low-voltage local
non-feedback electrical stimulation of monophasic and biphasic shapes. In
contrast to FitzHugh-Nagumo type models, this model accurately reproduces such
mesoscopic characteristics as action potential duration (APD), restitution and
conduction velocity (CV) found in cardiac tissue. It turns out that APD (CV) is a
function of both the previous APDs (CVs) and the time between excitations, also
known as the diastolic interval (DI) or recovery time \cite{ZJ}.

In our investigations we used a three variable simplified ionic model (SIM) of
the cardiac action potential, so--called Fenton-Karma equations
\cite{fentonkarma, fentonetal}:
\begin{equation}
\begin{array}{lll}
\partial_t u = \nabla(D\nabla u) - \left(J_{fi}(u,v) +
J_{so}(u) + J_{si}(u,w)\right),\\
\partial_t v = \Theta (u_c-u)(1-v)/\tau ^-_v(u) - \Theta (u-u_c)v/\tau
^+_v, \\
\partial_t w = \Theta (u_c-u)(1-w)/\tau ^-_w - \Theta (u-u_c)w/\tau
^+_w,
\end{array}
\label{eqFenton}
\end{equation}
where $u$ is a dimensionless membrane potential; $v,w$ are a fast
and a slow ionic gates, respectively; $D$ is a diffusion tensor
which is a diagonal matrix in our case (isotropic simulations);
$J_{fi}$, $J_{so}$, $J_{si}$ are scaled ionic currents
corresponding to the $Na$, $K$, $Ca$ currents, respectively;
$\Theta (x)$ is a standard Heaviside step function (other
functions, equations for the ionic currents and parameter values
see in \cite{fentonetal}). The parameters correspond to the steep
APD restitution (fitted to accurately represent the APD
restitution in the full Beeler-Reuter model) with breakup close to
the tip.

\begin{figure}[h]
\includegraphics[scale=0.8]{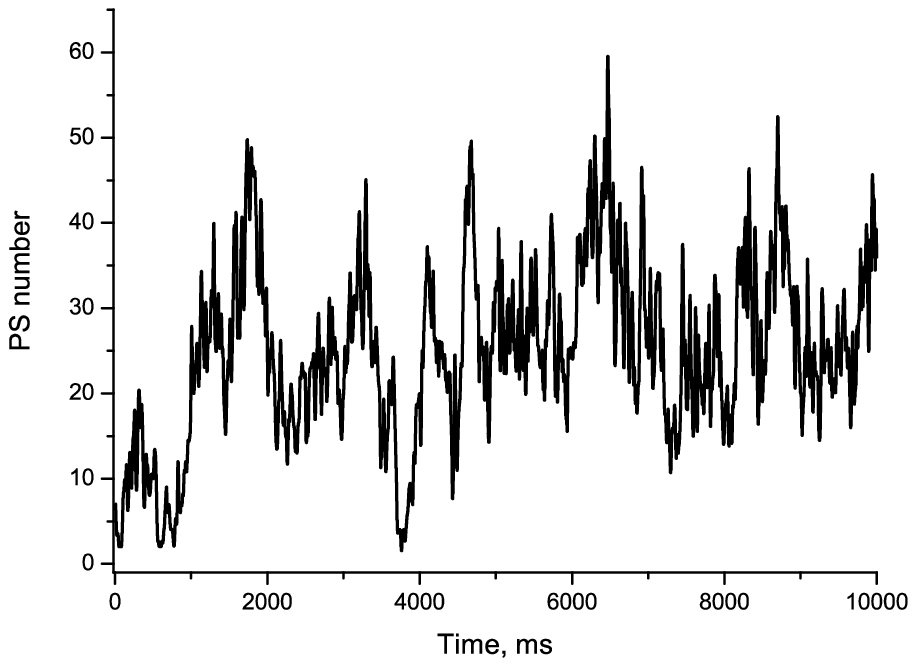}
\caption{The number of PSs as a function of time for the SIM with
parameters corresponding to the Set3 in the original Fenton-Karma
model \cite{fentonetal}.}\label{FilaCalc}
\end{figure}

The SIM supports many different mechanisms of spiral wave breakup
into complex re-entrant activity. Our numerical simulations were
performed in a 2D grid of $500\times500$ elements corresponding to
the tissue size of $12.5\times12.5$ cm. We used periodic boundary
conditions, which correspond to the torus topology. This geometry
is more close to a real geometry of ventricles than a sheet of
tissue (Neumann boundary conditions) and allowed us to exclude
wave attenuation. As a measure of the suppression of turbulent
effectiveness it is convenient to compute a number of phase
singularities (PSs), i.e. tips of re-entrant waves, which can be
detected by a number of various techniques \cite{zhuchkovaclayton,
clayzhuchpan}. We used the method described in \cite{bray}.

During numerical analysis we have found that an initial archimedean spiral wave
breaks into complex turbulent pattern after approximately 500 ms (see
Fig.\ref{FilaCalc}). Chaotic system states of 500, 600, 700 ms were considered as
initial ones for all our suppression attempts. To find the suppression effect, we
added the external almost {\it point} periodic forcing $J_{ext}(t)$ of the
frequency $\omega_{in}$ and the amplitude $A$ to the media. So, to stabilize the
spatio--temporal chaotic dynamics we generated in the medium a single pacemaker
(external electrode).

\begin{figure}[h]
\includegraphics[scale=0.4]{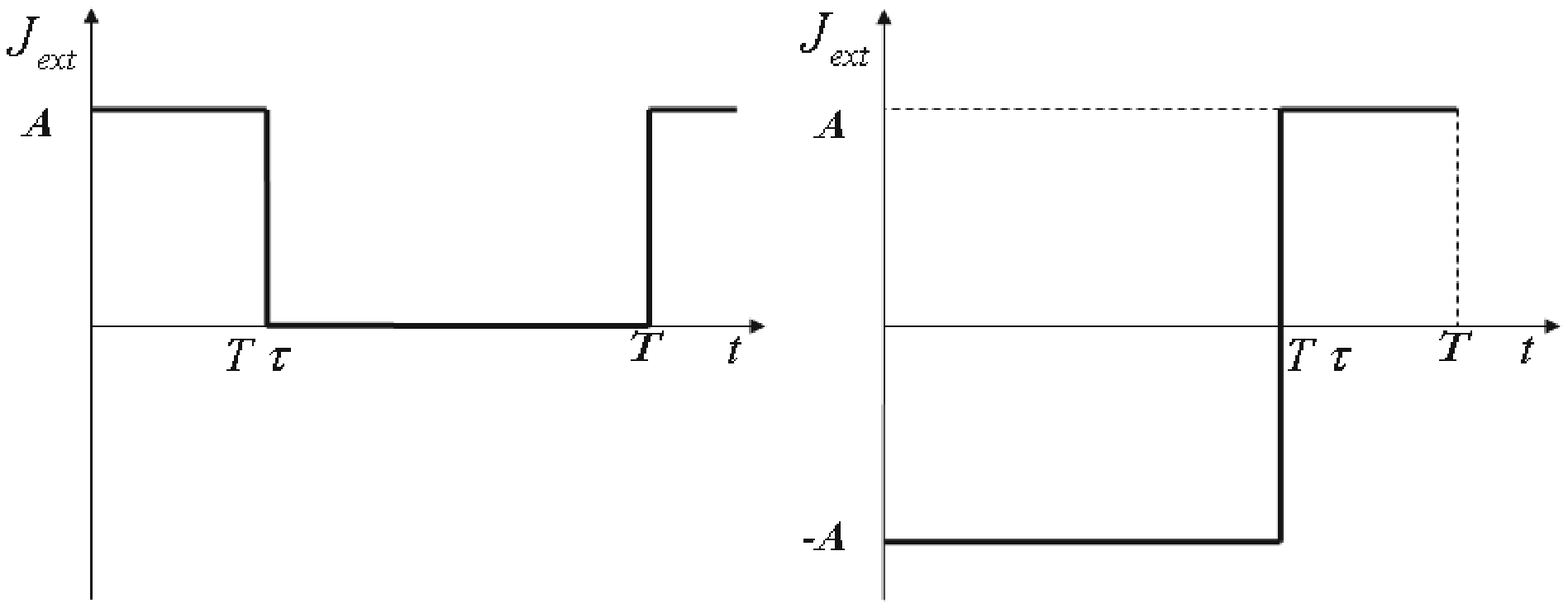}
\caption{Monophasic and biphasic impulses.} \label{shape}
\end{figure}

The shape of external stimulation $J_{ext}(t)$ is one of the key factors strongly
influencing on the suppression effectiveness. The defibrillation shocks used in
the clinical practise are of rectangular monophasic and biphasic shapes. In the
contrast to defibrillation by pulses (a single shock or series of shocks),
applied to the entire muscle or quite large part of it, we applied periodic
stimulation of the same mono(bi)phasic waveforms (Fig.\ref{shape}) to a point of
a medium ($2\times2$ nodes). Optimal values of $\tau$ ($\tau\leq1$) vary as
$0.05\div0.15$ for monophasic and $0.7\div0.75$ for biphasic stimuli. We took
$\tau = 0.1$ for the monophasic stimulation and $\tau = 0.7$ for the biphasic
waveform.

\begin{figure*}
\includegraphics[scale=0.8]{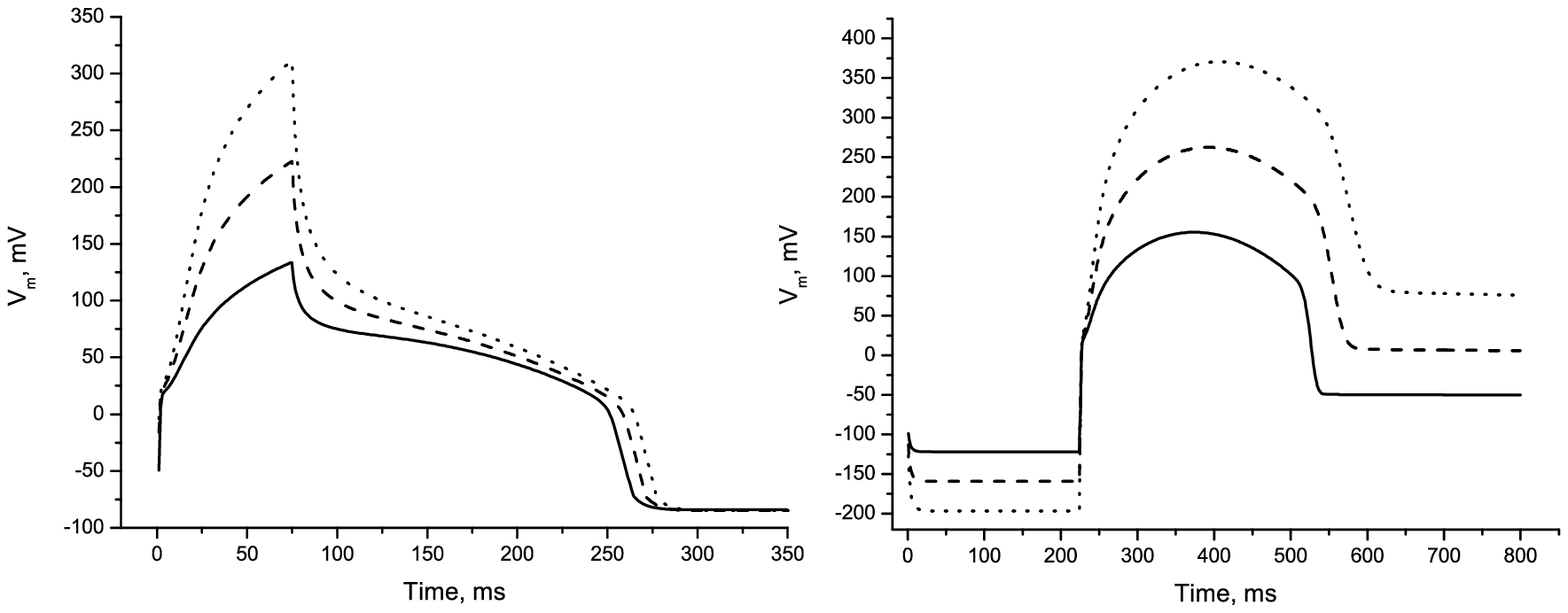}
\caption{\label{singlemonobi}Membrane potential at a pacemaker site vs time under
excitation by single stimuli with amplitudes $10, 20, 30{\mu}A/cm^{2}$ (bottom
up). Left--hand side --- monophasic stimulation, right--hand side --- biphasic
stimulation.}
\end{figure*}

\begin{figure*}
\includegraphics[scale=0.8]{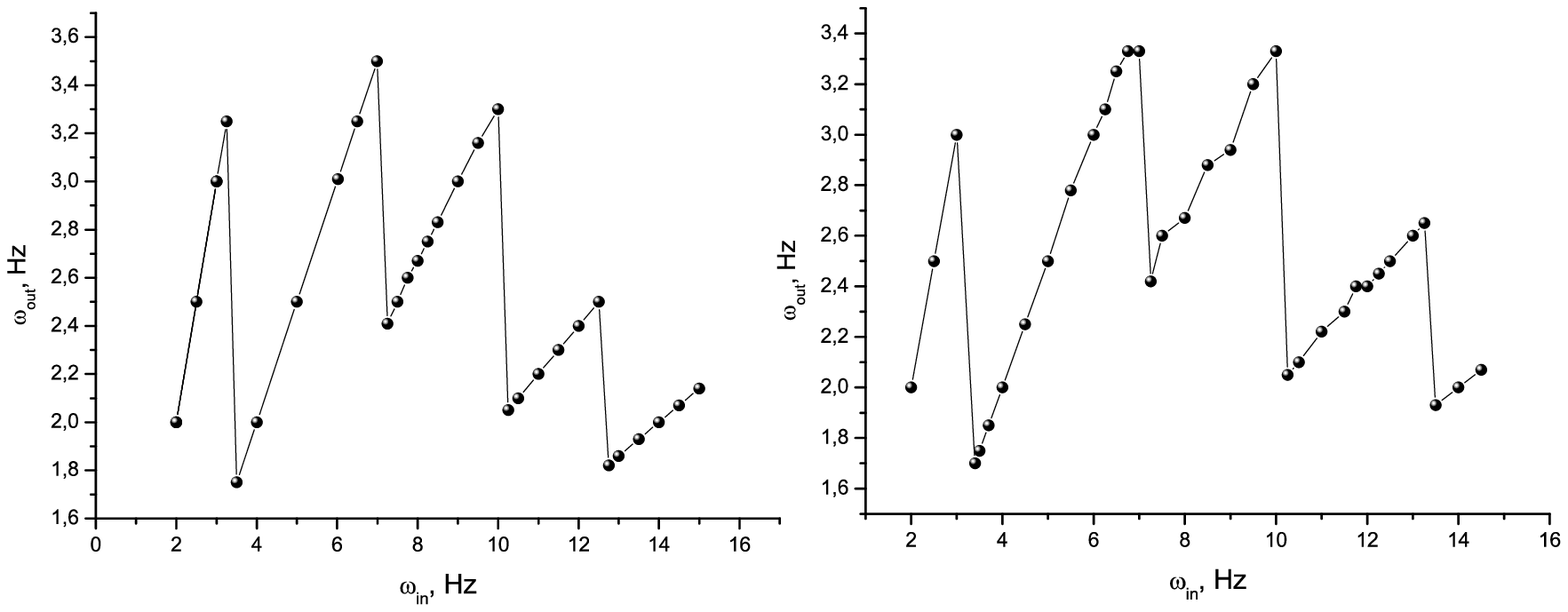}
\caption{\label{woutboth}The frequency $\omega_{out}$ of target waves as a
function of the pacemaker frequency $\omega_{in}$. Left--hand side --- monophasic
stimulation, right--hand side --- biphasic stimulation.}
\end{figure*}

\begin{figure}[h]
\includegraphics[scale=0.65]{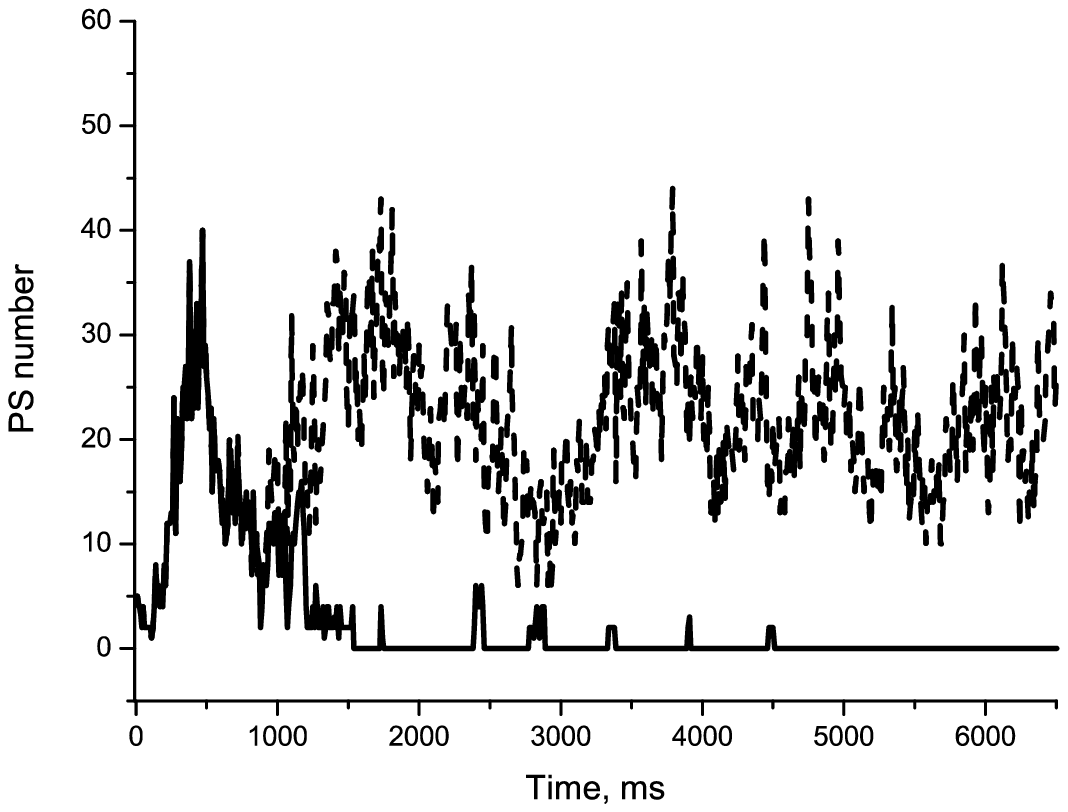}
\caption{\label{e025_mono_141ms_500600}The number of PSs as a function of time
for the SIM during monophasic stimulation of $\omega_{in}=7$ Hz and
$A=10{\mu}A/cm^2$. Suppression onset are: $t=500$ ms (solid lines), 600 ms
(dashes).}
\end{figure}

\begin{figure*}
\includegraphics[scale=0.8]{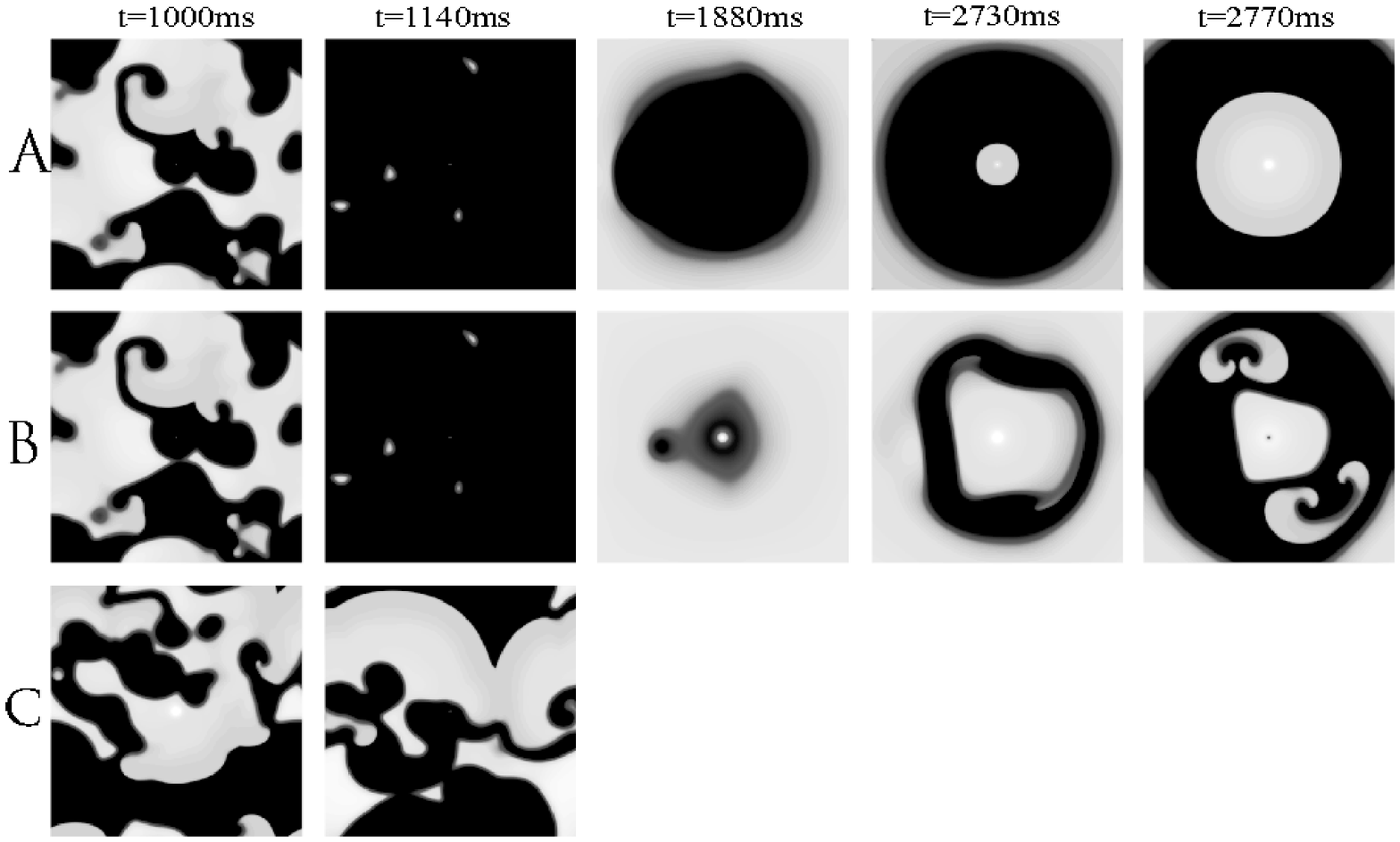}
\caption{\label{panels}Time evolution of the excitation pattern during biphasic
stimulation started at 600 ms with close frequencies. (A) Effective suppression
of complex activity by biphasic stimuli, $\omega_{in}=7.25$ Hz. At the subsequent
forcing all sites recovered without further activation. (B) Recovery of
spiral-wave activity after its suppression by biphasic stimuli, $\omega_{in}=7.0$
Hz. (C) Unsuccessful attempt to suppress turbulence by biphasic stimulation,
$\omega_{in}=6.75$ Hz.}
\end{figure*}

The important problem is to select stimulation amplitudes. To
determine the highest possible values of $A$ corresponding to
experimentally observable shock-induced variations of membrane
potential, we applied single stimuli of various amplitudes and
durations. For the monophasic waveform duration of influence,
$T\tau$ was 40 ms, which corresponds to $T = 400$ms, $\tau = 0.1$
for the periodic stimulation. For the biphasic stimulation
duration of the positive influence, $T(1-\tau)$ was 100 ms ($T =
333$ms, $\tau = 0.7$). Fig.\ref{singlemonobi} shows membrane
potential at a pacemaker site as a function of time for both
stimulation shapes and three amplitudes. To ensure experimentally
observed values of membrane potential for suppression we took $A =
10 {\mu}A/cm^{2}$ for both waveforms. Larger stimulation
amplitudes produce the positive shock-induced variations of
membrane potential larger than the maximal observed value of 100
mV \cite{fastrohrideker}. Because the cardiac membrane potential
varies as $-90\div +30$ mV, the maximal positive value of membrane
potential after applying shock is about 130 mV. Determination of
the lowest amplitude corresponding to the effective suppression is
a task of the future research.

We have measured the APD restitution curves obtained by two successive S1 and S2
stimuli of various amplitudes and durations (we use the 80\% cutoff ($APD_{80}$)
when calculating restitutions). It was found that the APD restitution curves are
the same for all stimuli and exactly look like one in Fig.4 in \cite{fentonetal}.
However, the S1-S2 interval of various stimuli is different to get the smallest
DI of 43 ms. It is less when stimulation amplitudes and/or durations are greater.
This means that the stimulation of larger amplitude and/or duration excites the
heart tissue being in relative refractory state earlier. This inverse dependence
of the refractory period on amplitude and duration of excitation phase does not
appear in FitzHugh-Nagumo type models.

Consider the problem of detecting the excitation frequencies, which provide the
effective stabilization. Since a search of the suppression frequencies at random
is ineffective, it is necessary to select the frequency of the stimulation close
to the maximal possible frequency for a given medium.

To find such frequencies, one can measure the period of target waves emitted by
the created source (electrode) as a function of its own period and then choose
values in the frequency intervals near the maxima of the frequency dependence. As
is known, if there are several co-existing sources of periodic waves in an
excitable medium, the interaction of waves leads to suppression of the sources
with a longer period by a source with a shorter period \cite{biktashev}. This was
firstly formulated in \cite{gelfandtsetlin} and is caused by the destructive
interaction of colliding waves in media, which mutually annihilate. If the
leading source is an external electrode, it can suppress re-entry subject to the
correct choice of its frequency and shape of stimulation.

Thus, to find the suppression frequencies we generated pacemakers in quite small
media volumes and determined the frequency $\omega_{out}$ of the target waves as
a function of the pacemaker frequency $\omega_{in}$ (Fig.\ref{woutboth}). One can
expect that the spatio-temporal chaos can be suppressed by the point external
perturbation in the frequency intervals corresponding to maxima of these
dependencies.

First, we tried to suppress complex activity by monophasic stimulation with
$\omega_{in}= 3.13,\ 7,\ 10$ Hz, corresponding to the frequency maxima on the
left--hand side of Fig.\ref{woutboth}. It was found that although the suppression
phenomenon was observed for $\omega_{in}=3.13$ Hz and $\omega_{in}=7$ Hz, it
strongly depends on the suppression onset. For example, the monophasic forcing of
$\omega_{in}=7$ Hz leads to the stabilization of chaotic dynamics if it starts at
500 ms, but suppression is unsuccessful if the suppression onset is 600 ms
(Fig.\ref{e025_mono_141ms_500600}). Vice versa, stimulation with
$\omega_{in}=3.13$ Hz started at 600 ms was successful. Suppression of turbulence
by stimuli with $\omega_{in}=10$ Hz failed for all considered three initial
conditions (500 ms, 600 ms, 700 ms). This phenomenon is connected with the
initial orientation of spiral tips with respect to the excitation source and
remans to be explored.

\begin{figure}[h]
\includegraphics[scale=0.8]{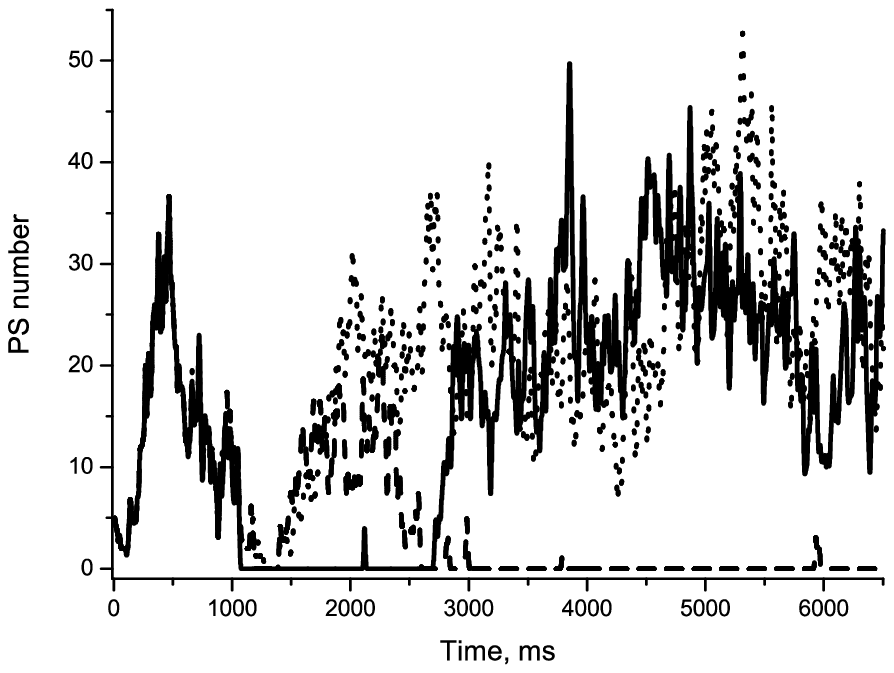}
\caption{The number of phase singularities as a function of time for the SIM
during biphasic stimulation of $\omega_{in}=7$ Hz, $A=10{\mu}A/cm^2$ (solid
lines), $20{\mu}A/cm^2$ (dashes), $30{\mu}A/cm^2$ (dots). The suppression onset
is $t=600$ ms.} \label{DifAmplSupBi}
\end{figure}

Second, we forced the system by the biphasic stimulation with $\omega_{in}=3.13,\
7.25,\ 10$ Hz corresponding to the frequency maxima on the right--hand side of
Fig.\ref{woutboth}. Again, the perturbation with $\omega_{in}=10$ Hz was
unsuccessful. But, in contrast to the monophasic stimulation, biphasic forcing
leads to the stabilization of complex dynamics by stimuli with $\omega_{in}=3.13$
Hz started at 500 ms and by stimulation with $\omega_{in}=7.25$ Hz started at 600
ms. Because the frequency interval corresponding to the second maximum on the
right--hand side of Fig.\ref{woutboth} is a quite wide, it is rather complicated
to select the appropriate value of the stimulation frequency. Fig. \ref{panels}
shows susceptibility to its choice. There is time evolution of the excitation
pattern during biphasic stimulation started at 600 ms with close frequencies 7.25
Hz (panel A), 7 Hz (panel B) and 6.75 Hz (panel C). The panel A corresponds to
the successful suppression (just an external pacemaker remained), B --- to the
recovered turbulence, C --- to the unsuccessful suppression. As it was predicted,
the patterns of the suppressed and recovered turbulence until suppressing
spiral-wave activity were similar due to the small difference in the stimulus
length, but creation of new spirals from an external pacemaker in the latter case
(panel B) was unexpected.

It should be noted that stimulation with $\omega_{in}=7$ Hz started at 600 ms
(resulting in recovered turbulence when $A=10{\mu}A/cm^2$) leads to the effective
suppression at doubling the stimulation amplitude. However, trebling the
amplitude is not helpful (Fig \ref{DifAmplSupBi}). So, there is a nonlinear
dependence on the stimulation amplitude.

Thus, although the suppression effectiveness strongly depends on the stimulation
frequency, amplitude and initial conditions (suppression onset), it was found
that for the correctly chosen these values the re-entrant waves can be easily
eliminated. It should be noted that the amplitude of the excitation is by three
orders of magnitude less than used in ICDs. Such low-voltage defibrillation has a
great advantage because it does not require knowledge of the re-entry frequency.
Moreover, in the case of VF, all the rotating waves are suppressed simultaneously
and the initial cardiac rhythm can be reestablished. By these reaosns, probably
the described new defibrillation strategy may be realizable in practice.

\end{document}